# V-VTEAM: A Compact Behavioral Model for Volatile Memristors

Tanay Patni, Rishona Daniels and Shahar Kvatinsky
*The Andrew and Erna Viterbi Faculty of Electrical and Computer Engineering*
*Technion – Israel Institute of Technology,* Haifa, Israel
Email: tanaypatni03@gmail.com , rishonad@campus.technion.ac.il, shahar@ee.technion.ac.il

*Abstract*—**Volatile memristors have recently gained popularity as promising devices for neuromorphic circuits, capable of mimicking the leaky function of neurons and offering advantages over capacitor-based circuits in terms of power dissipation and area. Additionally, volatile memristors are useful as selector devices and for hardware security circuits such as physical unclonable functions. To facilitate the design and simulation of circuits, a compact behavioral model is essential. This paper proposes V-VTEAM, a compact, simple, general, and flexible behavioral model for volatile memristors, inspired by the VTEAM nonvolatile memristor model and developed in MATLAB[1]. The validity of the model is demonstrated by fitting it to an ion drift/diffusion-based Ag/SiOx/C/W volatile memristor, achieving a relative root mean error square of 4.5%.**

## I. Introduction

Nonvolatile memristors have been quite popular for their applications as memory, in-memory computing, and hardware security [1]. Recently, there has been an increased interest in a different kind of memristor called a volatile memristor, as a potential device for neuromorphic computing. Nonvolatile memristors retain their state after the external signal is removed, while volatile memristors decay back to their initial state after the external signal is removed. Hence, volatile memristors can mimic the leaky nature of a neuron membrane, giving an advantage over nonvolatile memristors. Multiple neuron circuits using various devices like transistors, capacitors, and memristors have been proposed in the literature [2], and researchers are still actively trying to build better circuits. Neuron circuits designed by replacing transistors with memristors reduce power consumption and area and can better mimic the functioning of neuron membranes [3]. Volatile memristors are also used in circuits as selector devices and hardware security applications [4].

To further analyze and design circuits involving volatile memristors, it is essential to have a model to simulate and assess the performance of the circuit designs. In literature, very few volatile memristor models exist. The compact model published in [5] is physics-based and can closely emulate the specific device it is based on. However, simulations using this model

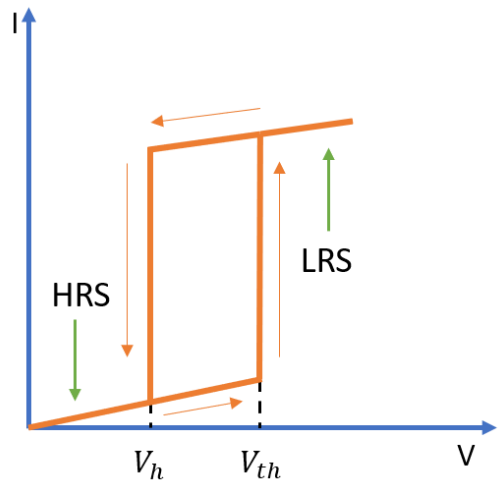
Figure 1: I-V curve of a typical volatile memristor

take considerable time and resources, and cannot simulate other types of volatile memristors. Thus, there is a need for a compact behavioral model focusing on flexibility. Hence, this paper proposes a compact, simple, general, and flexible behavioral volatile memristor model.

## II. Proposed Compact Model

### A. Volatile Memristor

A volatile memristor is a two-terminal device with a Metal-Insulator-Metal structure. It generally has two states: Low Resistance State (LRS) and High Resistance State (HRS). The typical I-V curve of a volatile memristor is shown in Fig. 1. Initially, the device is in HRS, and it changes its state to LRS when a voltage greater than the threshold voltage ($V_{th}$) of the device is applied across the terminals of the device. The device retains its LRS as long as the voltage applied across the device is greater than the hold voltage ($V_h$) . Once the applied voltage falls below $V_h$, the device switches back from LRS to HRS. The switching mechanism varies from device to device and is caused by electrochemical metallization or thermal chemical mechanism [4].

A memristive device is modeled using an internal variable that determines the state of the device [5]-[7]. The paper follows the same methodology and models the device with an internal

---

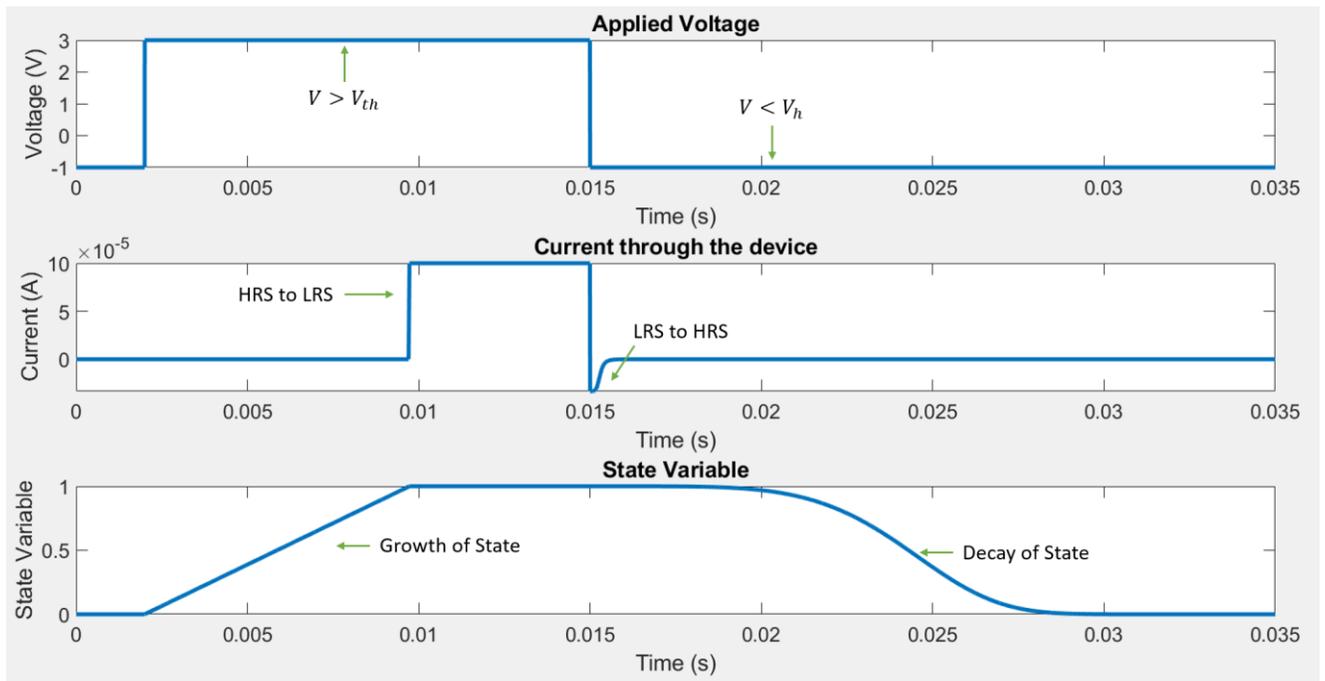

Figure 2: The applied voltage, current, and state variable of the proposed volatile memristor model.

state variable "$x$". The device is in HRS when "$x$" takes the minimum value "$x_{off}$" and in LRS when "$x$" takes the maximum value "$x_{on}$". The device switches state by the growth and the decay in the value of "$x$". The proposed model is unipolar but can easily be extended to bipolar by adding different parameter values to positive and negative applied voltages.

### B. Growth of Internal State Variable

When a voltage greater than $V_{th}$ is applied, the value of the internal state variable starts growing. This growth can be modeled by many different mathematical equations depending on the behavior of the device. The proposed model is inspired by VTEAM [8], a generic nonvolatile memristor model, with the following rate of growth of the internal state variable:

$$\frac{dx}{dt} = k \cdot \left(\frac{v(t)}{v_{th}} - 1\right)^{\alpha}, \qquad (1)$$

where k and $\alpha$ are constants, $v_{th}$ is the threshold voltage, and v(t) is the applied voltage. As shown in [8], (1) can fit many different devices and models with small errors; hence, it is used to model the change of the state variable.

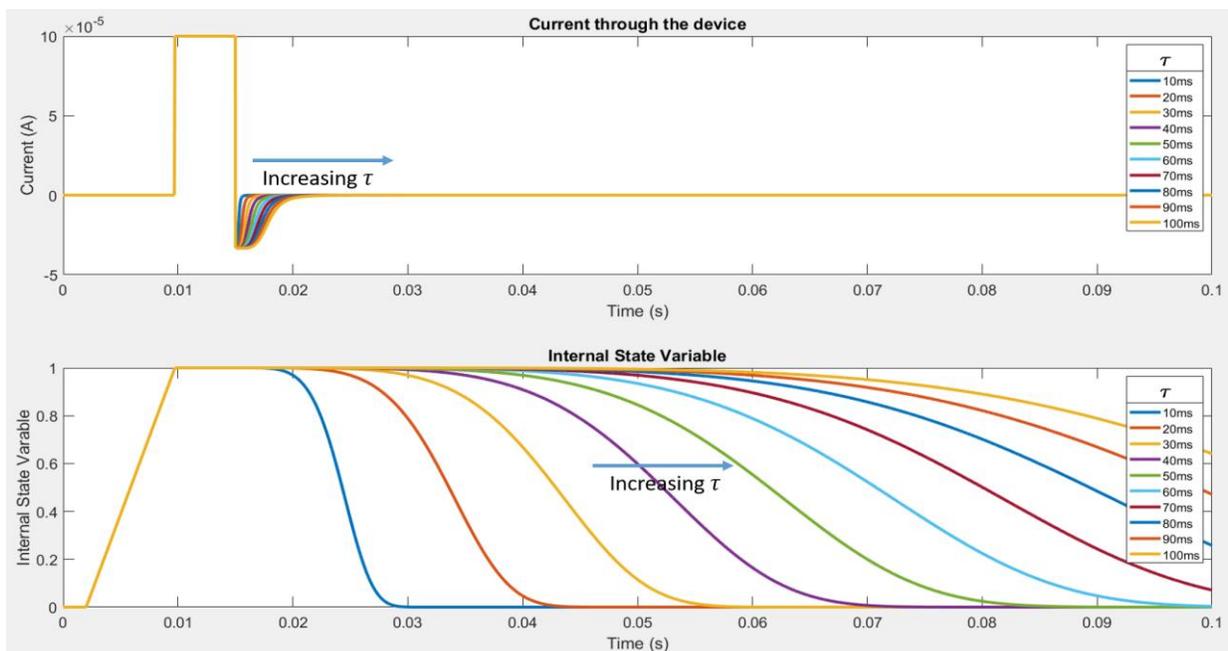

Figure 3: Retention time of the device for different $\tau$ values

## C. Decay of Internal State Variable

Unlike the growth of the internal state variable, the decay of the value is independent of the voltage applied once the applied voltage is lower than $V_h$. The value decays with time and is modeled using a stretched exponential function [9] given by

$$\frac{dx}{dt} = \frac{-x \cdot \beta \cdot \left(\frac{t}{\tau}\right)^{\beta-1}}{\tau}, \quad (2)$$

where $\beta$ and $\tau$ are the design variables. $\beta$ dictates the behavior of the decay and $\tau$ determines the time for the decay.

## D. Derivative Compact Model

The proposed compact model is the combination of the state derivative equations (1) and (2). The model is defined by

$$\frac{dx}{dt} = \begin{cases} \frac{-x \cdot \beta \cdot \left(\frac{t}{\tau}\right)^{\beta-1}}{\tau}, & v(t) \leq v_h \\ 0, & v_h < v(t) < v_{th} \\ k \cdot \left(\frac{v(t)}{v_{th}} - 1\right)^{\alpha}, & v(t) \geq v_{th} \end{cases} \quad (3)$$

## E. Electrical Modeling

To model the current-voltage relationship, a linear dependence of the resistance and internal state variable is assumed, similar to VTEAM [8], and the relationship is

$$i(t) = \left[R_{ON} + \frac{R_{OFF} - R_{ON}}{x_{off} - x_{on}} \cdot (x(t) - x_{on})\right]^{-1} \cdot v(t), \quad (4)$$

where $R_{on}$ and $R_{off}$ are the resistances of the device in LRS and HRS, respectively, and $x_{on}$ and $x_{off}$ are the respective values of the internal state variable.

## III. RESULTS

### A. Evaluation of the Model

A voltage signal is applied to the device to observe the model's functionality, and the current and the internal state variable are monitored. Fig. 2 shows the applied voltage, the current, and the internal state variable. A voltage of 3V, which is greater than the threshold voltage (1.8V), is applied to switch the state of the device from HRS to LRS, as seen by the increase in the device's current and the state variable. Once a voltage of -1V, lower than the hold voltage, is applied, a slow decay of the internal state variable is observed, which is also seen in the decrease of the current until it reaches close to zero. Note that while the internal state variable is growing, the current is not zero but very small. The parameter values used for this testing are listed in Table 1.

Retention time is one of the most important properties of a volatile memristor. It is the time taken by the device to switch from LRS to HRS once the applied voltage falls below $V_h$. As mentioned in Section II-C, the retention time for the model is determined by $\tau$. To observe the effect of $\tau$ on the functioning of the model, different values of $\tau$ from 10ms to 100ms with an interval of 10ms were evaluated for the same settings as the previous test. The results are shown in Fig. 3. It is observed that the retention time of the device increases with the increase in $\tau$.

TABLE I. Fitting Parameters of the model.

| Parameter | Testing Value | Fitting Value |
|---|---|---|
| $x_{on}$ | 1 | |
| $x_{off}$ | 0 | |
| $V_{th}$ [V] | 1.8 | 1.744 |
| $V_h$ [V] | 1.4 | 1.5726 |
| $R_{on}$ [Ω] | 30k | |
| $R_{off}$ [Ω] | 15G | |
| k [m/s] | 150 | 650 |
| $\alpha$ | 0.45 | 0.09999 |
| $\tau$ [s] | 10m | 1.0444 |
| $\beta$ | 5 | 2.14262 |

Note that $\tau$ is a parameter and not the actual retention time. The actual retention time is a function of both $\tau$ and $\beta$.

### B. Fitting to a Physics-Based Model

To test the flexibility and validity of the model, it was fitted to a physics-based model based on an ion drift/diffusion-based Ag/SiOx/C/W volatile memristor [5]. The error between the two models was determined using Relative Root Mean Square Error (RMSE) between the currents when the same input voltage signal was applied to both models. Relative RMSE is given by

$$e_i = \sqrt{\frac{1}{N}\left(\frac{\sum_{i=1}^{N}(I_{G,i} - I_{P,i})^2}{\bar{I_P^2}}\right)}, \quad (5)$$

where $N$ is the number of samples, $I_{G,i}$ and $I_{P,i}$ are the $i$-th samples of the generic and physics-based models, respectively. $\bar{I_P^2}$ is the Euclidian norm of the current of the physics-based model. The error was minimized using Gradient Descent [10] and simulated annealing algorithms [11]. The plot of the two models is shown in Fig. 4, and the parameter values are listed in Table 1. A relative RMSE of 4.5% was achieved.

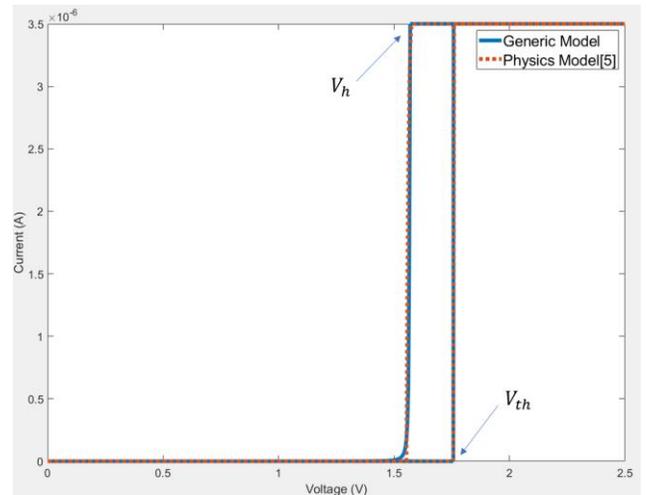

Figure 4: Proposed Generic Model fitted to an ion drift/diffusion-based Ag/SiOx/C/W volatile memristor model [5].

IV. CONCLUSION

With the growing interest in neuromorphic computing, the need for a generic compact volatile memristor model has become critical for designing circuits. This paper introduces a versatile and adaptable model aimed at meeting this need. The proposed model was implemented in MATLAB and has been validated by fitting a physics-based Ag/SiOx/C/W volatile memristor model, achieving a low relative RMSE, which demonstrates its accuracy and flexibility. This model can be adapted to various devices and enables rapid circuit simulations, facilitating the verification of novel circuits and supporting advancements in neuromorphic computing.

ACKNOWLEDGMENT

This project has received funding from the European Union's Horizon 2020 Research And Innovation Programme FET-Open NEU-Chip under grant agreement No. 964877.